\documentclass[12pt,a4paper]{article}

\usepackage[utf8]{inputenc}
\usepackage[T1]{fontenc}
\usepackage[english]{babel}
\usepackage{lmodern}
\usepackage{microtype}
\usepackage{setspace}

\usepackage{geometry}
\usepackage{amsmath,amssymb,amsfonts,bm}
\usepackage{graphicx}
\usepackage{booktabs}
\usepackage{array}
\usepackage{caption}
\usepackage{float}
\usepackage{flafter}
\usepackage{placeins}

\usepackage[round,authoryear]{natbib}
\usepackage[hidelinks]{hyperref}

\newcommand{\HIAR}{H-IAR}
\newcommand{\CIAR}{CIAR}
\newcommand{\BIAR}{BIAR}
\newcommand{\IAR}{IAR}
\newcommand{\Real}{\mathbb{R}}
\newcommand{\Quat}{\mathbb{H}}
\newcommand{\Phihat}{\widehat{\Phi}}
\newcommand{\norm}[1]{\left\lVert #1 \right\rVert}

\hypersetup{
  pdftitle={The H-IAR Model for Irregular Multispectral Time Series},
  pdfauthor={Bruno Goncalves C. Filho, Aluisio de Souza Pinheiro, Angelo Calil Bianchi},
  pdfsubject={Manuscript prepared for the Journal of Time Series Analysis},
  pdfkeywords={irregularly spaced time series, quaternion autoregression, state-space models, Kalman filtering, remote sensing, temporal persistence}
}

\begin{document}

\hypersetup{pageanchor=false}
\begin{titlepage}
\centering
\vspace*{2.0cm}
{\LARGE\bfseries The H-IAR Model for Irregular Multispectral Time Series\par}
\vspace{0.4em}
{\Large Quaternion Formulation, Mapping of Resilience Indicators, and Exploratory Identification of Forest Edges\par}

\vspace{2.0cm}
{\large Bruno Gon\c{c}alves C. Filho\textsuperscript{1}\par}
\vspace{0.35em}
{\large Angelo Calil Bianchi\textsuperscript{2}\par}
\vspace{0.35em}
{\large Alu\'isio de Souza Pinheiro\textsuperscript{3}\par}

\vspace{1.0cm}
{\normalsize
\textsuperscript{1}School of Mechanical Engineering (FEM),\\
University of Campinas (UNICAMP), Campinas, S\~ao Paulo, Brazil\par

\vspace{0.7em}
\textsuperscript{2}Institute of Science and Technology (ICT),\\
Federal University of S\~ao Paulo (UNIFESP),\\
S\~ao Jos\'e dos Campos, S\~ao Paulo, Brazil\par

\vspace{0.7em}
\textsuperscript{3}Department of Statistics,\\
Institute of Mathematics, Statistics and Scientific Computing (IMECC),\\
University of Campinas (UNICAMP), Campinas, S\~ao Paulo, Brazil\par
}

\vfill
\textbf{Corresponding author:} Aluísio de Souza Pinheiro\\
\textbf{Email for proofs:} \texttt{pinheiro@unicamp.br}
\end{titlepage}
\hypersetup{pageanchor=true}

\begin{abstract}
Satellite-based time series are often irregular because of clouds, shadows, atmospheric gaps, and orbital overlaps, while vegetation spectral responses involve correlated visible and near-infrared bands. We propose the Hypercomplex Irregular Autoregressive (\HIAR) model, a quaternion extension of the \IAR/\CIAR/\BIAR{} family for four-component observations at irregular times. Temporal dependence is represented by a real power of a quaternion parameter and estimated in state-space form through the Gaussian innovation likelihood of the Kalman filter, conditional on the adopted covariance specifications. Across 12,000 Monte Carlo fits, mean absolute bias decreased with sample size and ranged from 0.0009 to 0.0023 at $N=300$. We applied the model to 30,824 Sentinel-2 pixel series covering the Mata de Santa Genebra ARIE from 2020 to 2023. The optimizer reported successful numerical termination in 98.73\% of the fits; the median $\norm{\Phihat}$ was 0.6248, 3,872 pixels met the operational high-persistence threshold ($\norm{\Phihat}\geq0.95$), and the median residual RMSE for band B8 was 5.1551 percentage points. The results demonstrate the computational feasibility of \HIAR{} and its ability to generate descriptors of multispectral persistence, vector dynamics, predictive error, and spatial discontinuities potentially associated with forest edges.


\end{abstract}

\section{Introduction}

Assessing the resilience of forest ecosystems requires quantifying how environmental disturbances are absorbed and dissipated over time. In dynamical systems, proximity to a critical regime tends to reduce the recovery rate after perturbations, thereby increasing the temporal persistence of deviations from the mean state. This phenomenon, known as \textit{critical slowing down}, has been used as an indirect indicator of resilience loss and possible proximity to critical transitions in ecological systems \citep{scheffer2009,dakos2012}.

In remote sensing, this relationship has been investigated through autoregressive coefficients and temporal autocorrelations estimated from satellite vegetation series. Increasing temporal memory has been used as an indirect indicator of slower recovery and possible loss of forest resilience \citep{verbesselt2016,boulton2022,forzieri2022}. Such measures are nevertheless statistical descriptors that depend on the sensor, preprocessing choices, and observed variable; taken alone, they do not constitute direct measurements of ecological resilience \citep{smith2023}.

Vegetation reflectance in the visible and near-infrared regions can characterize leaf optical properties and canopy structure through remote sensing \citep{knipling1970,jacquemoud2009}. In particular, the Sentinel-2 constellation provides optical bands at a spatial resolution suitable for studying heterogeneous forest fragments \citep{esa2015}. Optical satellite time series, however, are rarely regularly spaced. Clouds, shadows, atmospheric variability, quality masks, and orbital overlaps create gaps and redundancies that preclude the direct use of classical discrete-time autoregressive models \citep{kandasamy2013,zhuwoodcock2014,rehfeld2011}.

A common solution is to interpolate or otherwise regularize the temporal grid before modeling. Although convenient, this approach may alter the autocorrelation structure and smooth the very residual component that carries information about temporal persistence \citep{rehfeld2011,kandasamy2013}. Irregular autoregressive models offer a more suitable alternative because they incorporate the observed time interval directly into the process dynamics \citep{eyheramendy2018,elorrieta2019}. The \IAR{} model was proposed for univariate observations collected at unequal time intervals \citep{eyheramendy2018}; the \CIAR{} model introduced a complex-valued representation that can be estimated with a Kalman filter \citep{elorrieta2019}; and the \BIAR{} model subsequently extended the approach to bivariate irregular time series, allowing two observed components to be modeled jointly \citep{elorrieta2021}.

Although the \IAR, \CIAR, and \BIAR{} models were developed and evaluated in studies strongly motivated by astronomical time series---particularly irregularly observed light curves and, in the bivariate case, different photometric bands \citep{eyheramendy2018,elorrieta2019,elorrieta2021}---their statistical structure is defined in terms of irregular temporal dependence rather than domain-specific physical properties. This feature suggests that the irregular-time core of the \IAR/\CIAR/\BIAR{} family can be transferred to environmental multispectral series, provided that the parameter interpretation is adapted to the new physical context.

Multispectral applications often involve more than two relevant bands and require a structure capable of representing marginal persistence and cross-component coupling within a single irregular dynamic system. Visible and near-infrared bands respond to distinct but potentially coupled optical and structural properties of vegetation \citep{knipling1970,jacquemoud2009}. Modeling each band separately ignores this coupling. An unrestricted vector transition would be more flexible, but it would not impose the parsimonious algebraic structure through which \HIAR{} combines radial contraction, associated with overall temporal persistence, with structured rotational mixing across components.

This paper introduces the Hypercomplex Irregular Autoregressive (\HIAR) model, a quaternion extension of the \IAR/\CIAR/\BIAR{} family. Four observed components are represented as a single element of the Hamilton quaternion algebra \citep{ward1997,kuipers1999}, and irregular temporal evolution is described through a real power of a quaternion parameter. This construction produces a real $4\times4$ transition matrix that is compatible with a state-space representation and estimation through the Gaussian innovation likelihood of the Kalman filter.

The main contributions are as follows:
\begin{enumerate}
    \item to formulate \HIAR{} as a quaternion autoregressive model for four-component irregularly spaced time series;
    \item to derive the real matrix representation of the transition $\Phi^{\Delta t}$ and embed it in a state-space model;
    \item to implement parameter estimation through the Gaussian innovation likelihood of the Kalman filter;
    \item to evaluate the finite-sample behavior of the estimator through Monte Carlo simulation; and
    \item to apply the model to Sentinel-2 series from the Mata de Santa Genebra ARIE, producing maps of temporal persistence, vector participation in the quaternion dynamics, predictive error, and spatial discontinuities potentially associated with forest edges.
\end{enumerate}

Table~\ref{tab:model_comparison} summarizes the conceptual progression of the irregular autoregressive family. Although \HIAR{} uses a $4\times4$ transition matrix, its structure is determined by a single quaternion parameter and is not an unrestricted vector transition.

\begin{table}[htbp]
\centering
\caption{Structural comparison of the precursor irregular autoregressive models and \HIAR.}
\label{tab:model_comparison}
\resizebox{\textwidth}{!}{%
\begin{tabular}{lllll}
\toprule
Model & Algebra & Observed components & Dynamic parameter & Main feature \\
\midrule
\IAR{} & $\Real$ & 1 & 1 real coefficient & Univariate irregular persistence \\
\CIAR{} & $\mathbb{C}$ & 1 plus one latent component & 1 complex coefficient & Oscillations and negative autocorrelations \\
\BIAR{} & $\mathbb{C}$ & 2 & 1 complex coefficient & Bivariate irregular dynamics \\
\HIAR{} & $\Quat$ & 4 & 1 quaternion & Structured contraction and rotation in $\Real^4$ \\
\bottomrule
\end{tabular}%
}
\end{table}

The remainder of the paper is organized as follows. Section~\ref{sec:model} presents the mathematical formulation of \HIAR. Section~\ref{sec:montecarlo} describes the Monte Carlo validation. Section~\ref{sec:data} details the satellite data and empirical pipeline. Section~\ref{sec:results} presents the mapping results. Section~\ref{sec:discussion} discusses the contributions, limitations, and methodological implications. Section~\ref{sec:conclusion} concludes.

\section{The H-IAR model}
\label{sec:model}

\subsection{Quaternion representation}

The \HIAR{} model is defined on the noncommutative real algebra of Hamilton quaternions, denoted by $\Quat$ \citep{ward1997,kuipers1999}. A quaternion $q\in\Quat$ can be written as
\[
    q=a+b\mathbf{i}+c\mathbf{j}+d\mathbf{k},
\]
where $a,b,c,d\in\Real$ and the imaginary units satisfy
\[
    \mathbf{i}^2=\mathbf{j}^2=\mathbf{k}^2=\mathbf{i}\mathbf{j}\mathbf{k}=-1.
\]
Multiplication is noncommutative and obeys the cyclic identities
$\mathbf{i}\mathbf{j}=\mathbf{k}$,
$\mathbf{j}\mathbf{k}=\mathbf{i}$, and
$\mathbf{k}\mathbf{i}=\mathbf{j}$, with a sign reversal when the order of the factors is exchanged.

For a four-component multispectral time series, each vector $X_{t_j}\in\Real^4$ is associated with the quaternion
\[
    q_{t_j}=x_{1,t_j}+x_{2,t_j}\mathbf{i}
             +x_{3,t_j}\mathbf{j}+x_{4,t_j}\mathbf{k}.
\]
Equivalently, define the vectorization map
\[
    \mathcal{V}:\Quat\longrightarrow\Real^4,
    \qquad
    \mathcal{V}\left(x_1+x_2\mathbf{i}+x_3\mathbf{j}+x_4\mathbf{k}\right)
    =
    \begin{bmatrix}
        x_1 & x_2 & x_3 & x_4
    \end{bmatrix}^{\top}.
\]
Thus, $X_{t_j}=\mathcal{V}(q_{t_j})$ and $\epsilon_{t_j}=\mathcal{V}(\varepsilon_{t_j})$. In the empirical application, the four components correspond, in order, to Sentinel-2 bands B2, B3, B4, and B8 after the preprocessing described in Section~\ref{sec:data}. Band B2 occupies the scalar component, whereas B3, B4, and B8 occupy the three imaginary components. This assignment is part of the adopted parameterization and is not invariant to arbitrary permutations that exchange scalar and vector roles.

\subsection{Irregular autoregressive equation}

Let $t_1<t_2<\cdots<t_N$ denote the observation times, and let $\Delta t_j=t_j-t_{j-1}$ be the interval between consecutive observations. The first-order \HIAR{} process is defined by
\begin{equation}
    q_{t_j}=\Phi^{\Delta t_j}q_{t_{j-1}}+\varepsilon_{t_j},
    \label{eq:hiar_quat}
\end{equation}
where $\Phi\in\Quat$ is the quaternion autoregressive parameter and $\varepsilon_{t_j}$ is a zero-mean quaternion disturbance. Dependence on $\Delta t_j$ allows the strength of temporal dependence to vary continuously with the observed interval, without interpolating the series onto a regular grid.

\subsection{Real powers of quaternions and the transition matrix}

To compute $\Phi^{\Delta t_j}$, we use the polar form of a nonzero quaternion \citep{kuipers1999,ward1997}. Let
\[
    \Phi=a+b\mathbf{i}+c\mathbf{j}+d\mathbf{k},
\]
with norm
\[
    \norm{\Phi}=\sqrt{a^2+b^2+c^2+d^2}.
\]
If the vector part $\mathbf{v}=(b,c,d)$ is nonzero, define the unit axis
\[
    \mathbf{u}=\frac{b\mathbf{i}+c\mathbf{j}+d\mathbf{k}}{\sqrt{b^2+c^2+d^2}}
\]
and the angle
\[
    \theta=\arccos\left(\frac{a}{\norm{\Phi}}\right).
\]
Under the principal polar branch,
\begin{equation}
    \Phi^{\Delta t_j}
    =
    \norm{\Phi}^{\Delta t_j}
    \left[
        \cos(\theta\Delta t_j)
        +
        \mathbf{u}\sin(\theta\Delta t_j)
    \right].
    \label{eq:polar_power}
\end{equation}
When $\sqrt{b^2+c^2+d^2}<10^{-12}$ and the scalar component is positive, the implementation uses the real limit $\Phi^{\Delta t_j}=a^{\Delta t_j}$. If $a<0$ and the vector part is zero, a real quaternion power has no unique limit independent of the approach axis when $\Delta t_j$ is noninteger. The implementation handles this numerically degenerate case through a zero-transition convention; this issue is revisited among the limitations.

Writing
\[
    \Phi^{\Delta t_j}=A_j+B_j\mathbf{i}+C_j\mathbf{j}+D_j\mathbf{k},
\]
left multiplication by $\Phi^{\Delta t_j}$, that is, the operator $q\mapsto\Phi^{\Delta t_j}q$, has the following real $4\times4$ matrix representation \citep{mebius2005,kuipers1999}:
\begin{equation}
F_{\Delta t_j}=
\left[
\begin{array}{rrrr}
A_j & -B_j & -C_j & -D_j \\
B_j &  A_j & -D_j &  C_j \\
C_j &  D_j &  A_j & -B_j \\
D_j & -C_j &  B_j &  A_j
\end{array}
\right].
\label{eq:transition_matrix}
\end{equation}
This matrix belongs to a structured subclass of linear transitions on $\Real^4$ and satisfies
\begin{equation}
    F_{\Delta t_j}^{\top}F_{\Delta t_j}
    =
    \norm{\Phi}^{2\Delta t_j}I_4.
    \label{eq:scaled_transition}
\end{equation}
Consequently, $F_{\Delta t_j}$ combines radial contraction by $\norm{\Phi}^{\Delta t_j}$ with a left-isoclinic rotation induced by the unit quaternion. Although it is a $4\times4$ matrix, it depends on only four components and is not a general vector transition. The resulting parsimony comes at the cost of restricting the admissible cross-component dependence patterns. Equation~\eqref{eq:hiar_quat} can therefore be written as a linear dynamic system on $\Real^4$:
\[
    X_{t_j}=F_{\Delta t_j}X_{t_{j-1}}+\epsilon_{t_j}.
\]

\subsection{State-space representation}

The state-space formulation separates the latent process dynamics from the observation equation \citep{kalman1960,durbin2012}:
\begin{equation}
\begin{aligned}
    X_{t_j}&=F_{\Delta t_j}X_{t_{j-1}}+\epsilon_{t_j},
    \qquad
    \epsilon_{t_j}\sim\mathcal{N}(0,Q_{t_j}),\\
    Y_{t_j}&=GX_{t_j}+e_{t_j},
    \qquad
    e_{t_j}\sim\mathcal{N}(0,R).
\end{aligned}
\label{eq:state_observation}
\end{equation}
All four components are observed directly in the multispectral application, so $G=I_4$. The matrix $R$ denotes the covariance of observational or instrumental error.

After centering the four components, the recursion is initialized with zero mean, $\widehat{X}_{t_1|t_0}=0$, and initial covariance $P_0$ equal to the diagonal matrix of empirical marginal variances. The first observation is neither assimilated through an update step nor included directly in the objective function, whose summation starts at $j=2$; it contributes only to the empirical construction of $P_0$.

\subsection{Dynamic-noise covariance}

Under an exact stationary formulation, let $P_0$ denote the target unconditional covariance. The dynamic-noise covariance should then satisfy the discrete Lyapunov equation \citep{durbin2012}
\begin{equation}
    Q_{t_j}=P_0-F_{\Delta t_j}P_0F_{\Delta t_j}^{\top}.
    \label{eq:lyapunov}
\end{equation}
This expression is straightforward when $P_0$ is proportional to the identity, because the rotation induced by $F_{\Delta t_j}$ preserves isotropy. A diagonal matrix is isotropic only when all marginal variances are equal, that is, $P_0=\sigma^2I_4$. Real multispectral bands may have substantially different variances. Imposing isotropy would remove this empirical anisotropy, whereas direct use of~\eqref{eq:lyapunov} with anisotropic $P_0$ can produce $Q_{t_j}$ matrices that are not positive semidefinite while the numerical optimizer explores the parameter space.

In the implementation, $P_0$ is the diagonal matrix of empirical variances of the four centered components. This preserves the marginal scale of each band without estimating additional cross-covariances. We adopt the following approximate diagonal specification:
\begin{equation}
    Q_{t_j}
    =
    P_0\left(1-\norm{\Phi}^{2\Delta t_j}\right).
    \label{eq:q_approx}
\end{equation}
This choice preserves the empirical marginal variances on the diagonal of $P_0$, keeps $Q_{t_j}$ positive semidefinite for $\norm{\Phi}<1$, and stabilizes large-scale estimation. In the isotropic case $P_0=\sigma^2I_4$, it coincides with the condition that preserves the unconditional covariance because the rotational part of $F_{\Delta t_j}$ preserves scalar matrices. In the anisotropic case, it is a first-order operational approximation: $P_0$ acts as the initial covariance and as a reference matrix of marginal variances, but is not exactly preserved as the unconditional covariance under the approximate dynamics. The diagonal approximation for $Q_{t_j}$ does not force predicted or filtered covariances to remain diagonal; the full Kalman recursion can produce cross-covariances through $F_{\Delta t_j}$.

\subsection{Kalman filter and innovation likelihood}

Given the state-space formulation in~\eqref{eq:state_observation}, the Kalman innovations are computed recursively \citep{kalman1960,durbin2012,elorrieta2019}. The prediction step is
\begin{align*}
    \widehat{X}_{t_j|t_{j-1}}
    &=F_{\Delta t_j}\widehat{X}_{t_{j-1}|t_{j-1}},\\
    P_{t_j|t_{j-1}}
    &=F_{\Delta t_j}P_{t_{j-1}|t_{j-1}}F_{\Delta t_j}^{\top}+Q_{t_j}.
\end{align*}
The innovation and its covariance are
\begin{align*}
    v_{t_j}&=Y_{t_j}-G\widehat{X}_{t_j|t_{j-1}},\\
    \Lambda_{t_j}&=GP_{t_j|t_{j-1}}G^{\top}+R.
\end{align*}
The Kalman gain is
\[
    K_{t_j}=P_{t_j|t_{j-1}}G^{\top}\Lambda_{t_j}^{-1},
\]
and the update is
\begin{align*}
    \widehat{X}_{t_j|t_j}
    &=\widehat{X}_{t_j|t_{j-1}}+K_{t_j}v_{t_j},\\
    P_{t_j|t_j}
    &=(I_4-K_{t_j}G)P_{t_j|t_{j-1}}.
\end{align*}

When the innovation covariance is critically ill-conditioned, a small diagonal term $\gamma I_4$ is added before computing its determinant and inverse. The implementation uses $\gamma=10^{-6}$. This term acts as diagonal numerical jitter only in such cases and is not estimated as an additional component of the observation model.

Conditional on $P_0$, obtained by substituting the empirical variances, and on $R$, fixed for each application, the objective function is the negative Gaussian innovation log-likelihood \citep{durbin2012,elorrieta2019}:
\begin{equation}
    \mathcal{L}(\Phi)
    =
    \frac{1}{2}
    \sum_{j=2}^{N}
    \left[
        k\log(2\pi)
        +
        \log|\Lambda_{t_j}|
        +
        v_{t_j}^{\top}\Lambda_{t_j}^{-1}v_{t_j}
    \right],
    \label{eq:nll}
\end{equation}
with $k=4$ in the full application. This is therefore a Gaussian innovation likelihood with covariance parameters specified by a plug-in procedure, rather than a joint estimation of all components of $P_0$, $Q_{t_j}$, and $R$. Equation~\eqref{eq:nll} is minimized with L-BFGS-B \citep{byrd1995,zhu1997}. Componentwise bounds restrict the search to a compact region, while a radial penalty imposes
\[
    \norm{\Phi}^2\leq0.99.
\]
Whenever a proposed point exceeds this radial limit, the objective is evaluated at its radial projection onto the admissible boundary plus a quadratic penalty proportional to the excess.

L-BFGS-B was run with a maximum of 2,000 iterations and relative objective tolerance \texttt{ftol}$=10^{-9}$. A fit is classified as numerically successful when the returned optimization object has \texttt{success=True}. This flag certifies only that a numerical stopping criterion was met; it does not guarantee a global optimum or statistical adequacy.

\section{Monte Carlo validation}
\label{sec:montecarlo}

Monte Carlo experiments were used to assess the finite-sample behavior, numerical stability, and computational scaling of the \HIAR{} estimator \citep{eyheramendy2018,elorrieta2019,elorrieta2021}. The purpose is not to provide an analytical proof of consistency, but to determine empirically whether the algorithm recovers the true parameters in synthetic series generated from the model.

\subsection{Generation of irregular observation times}

To reproduce unequally spaced observation patterns, the intervals $\Delta t_j$ were generated from a mixture of two exponential distributions. Because the implementation uses the scale parameter of the exponential distribution, the density is
\[
    f(\Delta t)
    =
    w_1\frac{1}{\beta_1}\exp\left(-\frac{\Delta t}{\beta_1}\right)
    +
    w_2\frac{1}{\beta_2}\exp\left(-\frac{\Delta t}{\beta_2}\right),
\]
where $\beta_1=15$, $\beta_2=2$, $w_1=0.15$, and $w_2=0.85$. Observation times are obtained by cumulative summation of these intervals, with $t_1=0$.

\subsection{Parameter scenarios}

Four values of $\Phi$ were considered, combining positive and negative signs in the scalar and vector components:
\begin{align*}
    \Phi_1 &=  0.70 + 0.30\mathbf{i} + 0.30\mathbf{j} + 0.30\mathbf{k}, \\
    \Phi_2 &= -0.70 - 0.30\mathbf{i} - 0.30\mathbf{j} - 0.30\mathbf{k}, \\
    \Phi_3 &= -0.90 + 0.15\mathbf{i} + 0.15\mathbf{j} + 0.15\mathbf{k}, \\
    \Phi_4 &=  0.90 - 0.15\mathbf{i} - 0.15\mathbf{j} - 0.15\mathbf{k}.
\end{align*}
For each scenario, $M=1{,}000$ independent replications were generated at sample sizes
\[
    N\in\{30,100,300\}.
\]
Synthetic series were simulated with $P_0=I_4$, initial state $X_{t_1}\sim\mathcal{N}(0,I_4)$, and dynamic noise
\[
    Q_{t_j}=I_4\left(1-\norm{\Phi}^{2\Delta t_j}\right).
\]
Under this isotropic data-generating mechanism, the expression for $Q_{t_j}$ coincides exactly with the stationary covariance condition. For each fitted sample, however, the estimator recomputed $P_0$ as the diagonal matrix of empirical variances and used $R=10^{-6}I_4$ when no observation covariance was supplied. The validation therefore also incorporates the plug-in procedure used in estimation. Moreover, the scenarios with a negative scalar component have nonzero vector parts and do not fall on the degenerate negative real semiaxis described in Section~\ref{sec:model}.

\subsection{Evaluation criteria}

For each component $\ell\in\{a,b,c,d\}$, we computed the mean estimate, empirical absolute bias, and Monte Carlo standard deviation:
\begin{align*}
    \overline{\widehat{\Phi}}_{\ell}
    &=\frac{1}{M}\sum_{m=1}^{M}\widehat{\Phi}_{\ell}^{(m)},\\
    |\mathrm{Bias}_{\ell}|
    &=\left|\overline{\widehat{\Phi}}_{\ell}-\Phi_{\ell}\right|,\\
    SD_{\ell}
    &=
    \sqrt{
    \frac{1}{M-1}
    \sum_{m=1}^{M}
    \left(
        \widehat{\Phi}_{\ell}^{(m)}-
        \overline{\widehat{\Phi}}_{\ell}
    \right)^2
    }.
\end{align*}
Mean estimation time and mean optimizer iteration count were also recorded.

\begin{table}[htbp]
\centering
\caption{Aggregate Monte Carlo results for the \HIAR{} estimator. Bias and dispersion metrics are aggregated across the four components $a,b,c,d$ within each scenario and sample size.}
\label{tab:montecarlo_summary}
\resizebox{\textwidth}{!}{%
\begin{tabular}{llrrrrrr}
\toprule
Case & $N$ & Mean abs. bias & Max. abs. bias & Mean SD & Iterations & Mean time (s) & Time/iter. (ms) \\
\midrule
1 (+/+) & 30  & 0.0150 & 0.0212 & 0.0572 & 10.663 & 0.0153 & 1.4321 \\
1 (+/+) & 100 & 0.0048 & 0.0064 & 0.0279 &  9.955 & 0.0291 & 2.9243 \\
1 (+/+) & 300 & 0.0014 & 0.0024 & 0.0153 &  9.384 & 0.0814 & 8.6776 \\
2 (-/-) & 30  & 0.0219 & 0.0360 & 0.0557 & 13.177 & 0.0137 & 1.0407 \\
2 (-/-) & 100 & 0.0058 & 0.0102 & 0.0213 & 12.146 & 0.0376 & 3.0955 \\
2 (-/-) & 300 & 0.0023 & 0.0036 & 0.0120 & 11.413 & 0.1048 & 9.1810 \\
3 (-/+) & 30  & 0.0220 & 0.0579 & 0.0645 & 16.736 & 0.0197 & 1.1785 \\
3 (-/+) & 100 & 0.0044 & 0.0123 & 0.0108 & 16.827 & 0.0617 & 3.6640 \\
3 (-/+) & 300 & 0.0015 & 0.0041 & 0.0059 & 17.580 & 0.1929 & 10.9751 \\
4 (+/-) & 30  & 0.0080 & 0.0215 & 0.0325 & 13.972 & 0.0144 & 1.0322 \\
4 (+/-) & 100 & 0.0024 & 0.0049 & 0.0150 & 14.188 & 0.0427 & 3.0108 \\
4 (+/-) & 300 & 0.0009 & 0.0019 & 0.0085 & 13.594 & 0.1239 & 9.1163 \\
\bottomrule
\end{tabular}%
}
\end{table}

\begin{figure}[htbp]
    \centering
    \includegraphics[width=0.85\textwidth]{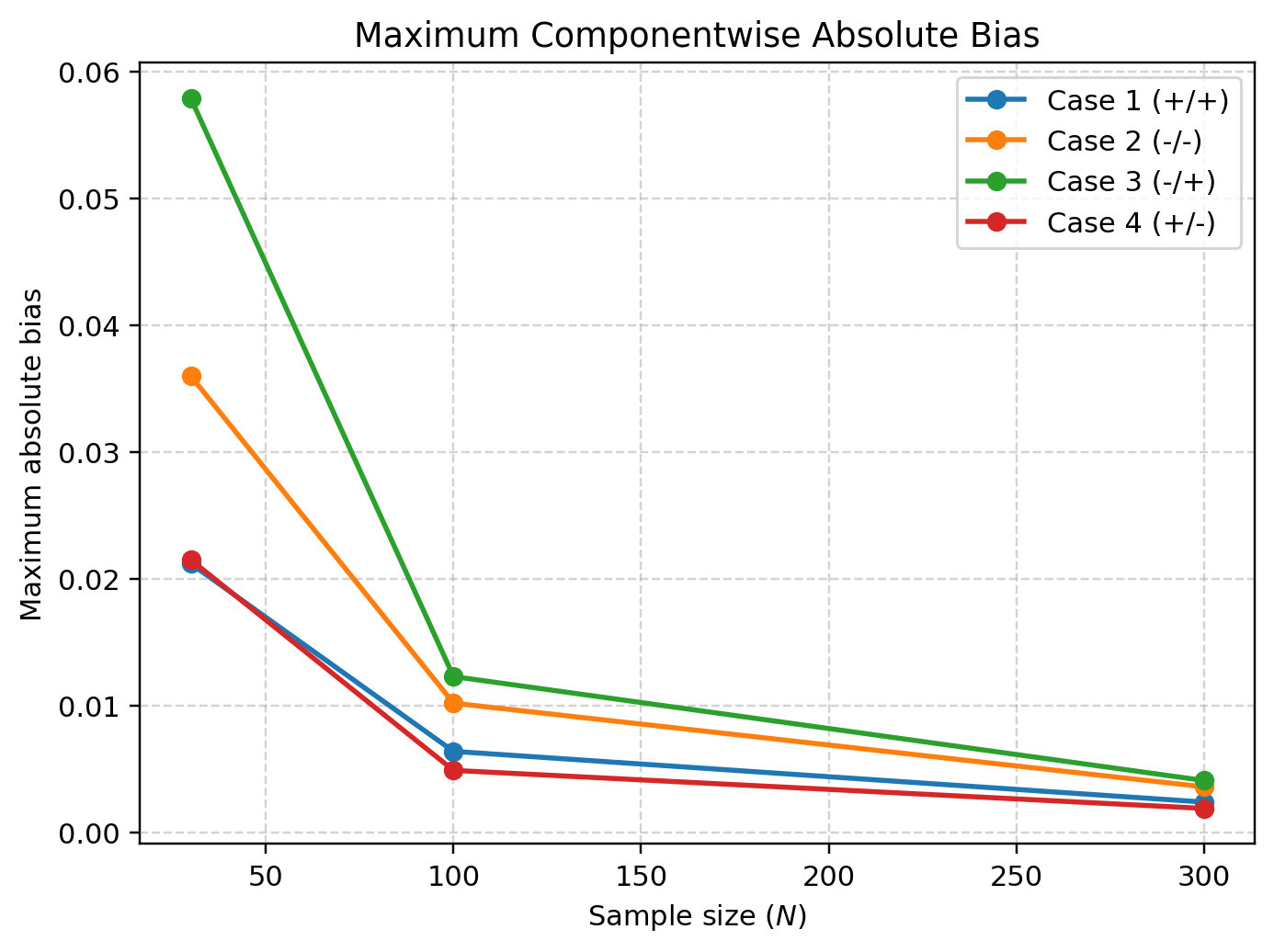}
    \caption{Maximum componentwise absolute bias of the \HIAR{} estimator in the Monte Carlo experiments, by sample size and parameter scenario.}
    \label{fig:mc_bias}
\end{figure}

\begin{figure}[htbp]
    \centering
    \includegraphics[width=0.85\textwidth]{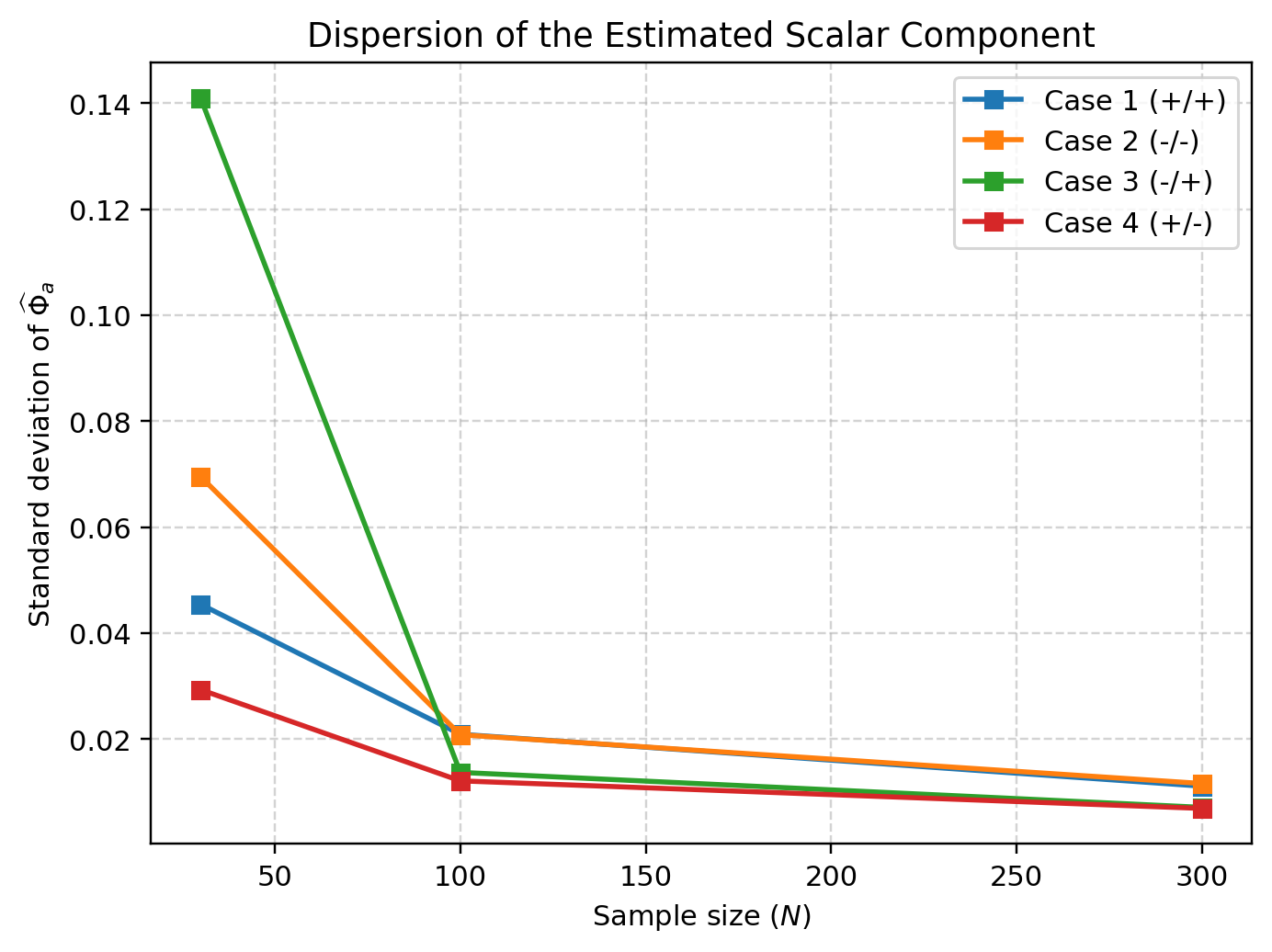}
    \caption{Monte Carlo standard deviation of the estimated scalar component $\widehat{\Phi}_a$, by sample size and parameter scenario.}
    \label{fig:mc_sd}
\end{figure}

\begin{figure}[htbp]
    \centering
    \includegraphics[width=0.85\textwidth]{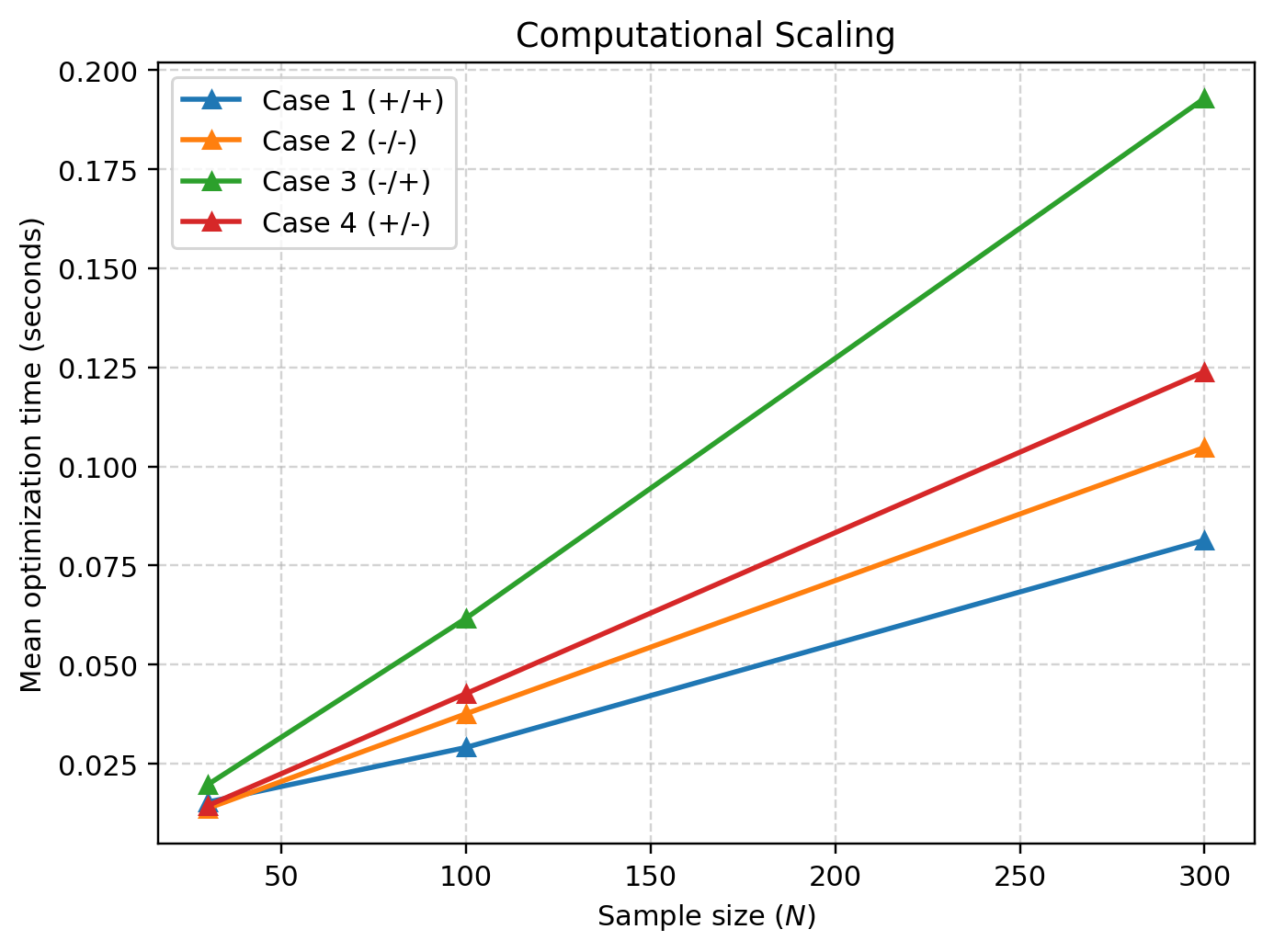}
    \caption{Mean \HIAR{} estimation time in the Monte Carlo experiments, by sample size and parameter scenario.}
    \label{fig:mc_time}
\end{figure}

Figures~\ref{fig:mc_bias} and~\ref{fig:mc_sd} show that both the maximum componentwise absolute bias and the empirical dispersion of the scalar estimate decrease with $N$ in all four scenarios. Together with the full componentwise results in Appendix~\ref{app:montecarlo}, these patterns provide numerical evidence of estimator stability and parameter recovery. Figure~\ref{fig:mc_time} reports the mean estimation cost as a function of sample size, with growth compatible with the sequential Kalman-filter recursion. These numerical patterns do not constitute a formal asymptotic proof.

\section{Satellite data and empirical pipeline}
\label{sec:data}

\subsection{Study area}

The empirical application was conducted in the Mata de Santa Genebra Area of Relevant Ecological Interest (ARIE), located in Campinas, S\~ao Paulo, Brazil \citep{icmbio_santa_genebra,guirao2011}. The site is an ecologically relevant forest fragment embedded in an urban matrix \citep{guirao2011,icmbio_santa_genebra}, which permits the joint examination of forest interior, edges, and internal heterogeneity.

\subsection{Sentinel-2 acquisition and Google Earth Engine extraction}

Time series were extracted through the Google Earth Engine API \citep{gorelick2017} from the \texttt{COPERNICUS/S2\_SR\_HARMONIZED} collection, which provides harmonized Sentinel-2 Level-2A surface reflectance \citep{gee_s2}. This collection was selected to reduce radiometric discontinuities associated with Processing Baseline 04.00 and to improve comparability between scenes acquired before and after 2022 \citep{gee_s2}.

The Phase B extraction region was defined by a closed polygon covering the forest core and its peripheral fringe. In each scene, the Scene Classification Layer (SCL) was used to retain only pixels classified as vegetation, that is, \texttt{SCL == 4} \citep{gee_s2}. Bands B2, B3, B4, and B8 were then selected as the four multispectral components of the quaternion vector \citep{esa2015,gee_s2}. Digital values were divided by 100, converting the scaled digital numbers into percentage reflectance \citep{gee_s2}.

The \texttt{system:time\_start} timestamp was inserted as a constant band named \texttt{time}, ensuring that each exported sample contained the scene time in addition to coordinates and spectral bands. Spatial sampling used a 10-m resolution \citep{esa2015,gee_s2}, and the resulting spatiotemporal table was asynchronously exported to Google Drive in CSV format.

\subsection{Phase A: pilot validation}

Before regional processing, Phase A tested the pipeline on a controlled pilot sample of at most 30 pixels from Mata de Santa Genebra. This stage validated Sentinel-2 extraction, vegetation filtering, time conversion, within-day aggregation, removal of deterministic components, and the call to the \HIAR{} estimator.

For each pixel, absolute time in milliseconds was converted to days since January 1, 2020. Observations from the same day were averaged to avoid zero or near-zero $\Delta t$ values. Pixels with fewer than 30 valid observations were discarded. The remaining series were fitted with a deterministic model containing an annual harmonic component and a linear trend \citep{wilson2018,verbesselt2010}:
\begin{equation}
    y(t)
    =
    \beta_0
    +\beta_1\cos(\omega t)
    +\beta_2\sin(\omega t)
    +\beta_3t,
    \qquad
    \omega=\frac{2\pi}{365.25}.
    \label{eq:fourier}
\end{equation}
Residuals were centered componentwise and passed to the \HIAR{} estimator so that autoregressive dynamics were estimated around the residual mean of each band. Phase A produced a pixel-level table containing geographic coordinates, the number of valid observations, the estimated quaternion, the global norm $\norm{\Phihat}$, the numerical-success flag, and the optimizer iteration count.

\subsection{Phase B: regional processing}

In Phase B, the full spatiotemporal table was processed by unique coordinate pairs. After within-day consolidation and chronological splitting, pixels with fewer than 30 training observations were discarded, leaving 30,824 valid pixels. Each pixel was treated as an independent time series. Multicore CPU parallelization transformed the pointwise estimator into a spatial mapping pipeline. Processing was performed on the shared Zurich server at IMECC/UNICAMP, which has 24 CPU cores and 128 GB of RAM. Under these conditions, the full grid was processed in approximately 10 minutes. This value is an operational record for the reported execution, not a hardware-independent benchmark.

The observation-error covariance was fixed at
\[
    R=4I_4,
\]
as an operational specification chosen \textit{a priori}. Because the bands were represented in percentage-reflectance points, the diagonal value $4=2^2$ corresponds to an observational standard deviation of 2 percentage points per band. The structure $4I_4$ assumes a common observation-error variance across the four components, no contemporaneous correlation among measurement errors, and constant uncertainty across times and pixels. The value was neither estimated from the data nor subjected to a sensitivity analysis; the empirical results are therefore conditional on this specification.

To prevent data leakage, each pixel series was split chronologically into 90\% for training and 10\% for validation. The deterministic model in~\eqref{eq:fourier} was fitted exclusively to the training segment. Its estimated parameters were then used to remove seasonality and trend from both training and test observations. If the harmonic fit did not converge, a reduced deterministic adjustment based only on the training-set mean was used. Thus, no future information entered the deterministic fit. During estimation, the \HIAR{} routine internally centered the training residuals componentwise, preserving the zero-mean latent-process formulation.

After estimating $\Phihat$ on the training set, predictive validation was performed on the multispectral residuals in the test set. The first test prediction started from the last observed training residual. Each subsequent prediction used the residual actually observed at the immediately preceding test time. This is conditional one-step-ahead validation rather than free-running multi-step forecasting. The rule propagates the previous observed residual directly through $F_{\Delta t_j}$, not the filtered latent state; when $R\neq0$, it differs from the predictive mean of the full Kalman recursion. Squared errors were computed between predicted and observed residuals on the percentage-reflectance scale. Band-specific root mean squared error (RMSE) was used as the local predictive-performance measure:
\[
    RMSE_b
    =
    \sqrt{
    \frac{1}{n_{\mathrm{test}}}
    \sum_{i=1}^{n_{\mathrm{test}}}
    \left(
        x^{\mathrm{res}}_{b,i}-\widehat{x}^{\mathrm{res}}_{b,i}
    \right)^2
    }.
\]

\subsection{Rasterization and derived maps}

Pixel-level estimates were converted into georeferenced GeoTIFF surfaces. Coordinates were mapped directly to raster indices without continuous spatial interpolation, thereby avoiding synthetic values in clearings, indentations, and regions lacking validated vegetation observations.

The first derived surface was multispectral temporal persistence, operationally defined as the Euclidean norm of the quaternion parameter:
\begin{equation}
    \mathcal{P}
    =\norm{\Phihat}
    =
    \sqrt{
        \widehat{\Phi}_a^2+
        \widehat{\Phi}_b^2+
        \widehat{\Phi}_c^2+
        \widehat{\Phi}_d^2
    }.
    \label{eq:persistence}
\end{equation}
Because time was measured in days, $\mathcal{P}$ is the reference persistence factor for a one-day interval, while contraction over an interval of $h$ days is $\mathcal{P}^{h}$. By~\eqref{eq:scaled_transition}, larger $\mathcal{P}$ values imply stronger retention of multispectral deviations and slower dissipation, whereas smaller values imply faster dissipation. Within the \textit{critical slowing down} paradigm, high persistence is compatible with slower recovery and possible resilience loss, not with greater resilience in a direct sense.

The second surface was vector dominance, defined as
\begin{equation}
    \mathcal{D}
    =
    \frac{
        |\widehat{\Phi}_b|+
        |\widehat{\Phi}_c|+
        |\widehat{\Phi}_d|
    }{
        |\widehat{\Phi}_a|+10^{-8}
    }.
    \label{eq:vector_dominance}
\end{equation}
This quantity summarizes the relative contribution of the vector components compared with the scalar component.

The metric $\mathcal{D}$ was proposed in this study as an exploratory descriptor and is not a standardized index from the literature. The term $10^{-8}$ serves only to prevent division by zero. Because $\mathcal{D}$ is unbounded, it can become large when $|\widehat{\Phi}_a|$ approaches zero even if the absolute magnitude of the vector part is modest. Furthermore, the $L^1$ sum depends on the representation chosen for the imaginary components. The index should therefore be interpreted comparatively within the analyzed domain.

The third surface was predictive RMSE, with particular attention to near-infrared band B8. The fourth was the morphological gradient of persistence. To compute it, missing cells in the persistence raster were temporarily filled by nearest-neighbor values solely to stabilize convolution and prevent propagation of invalid values. The Sobel operator was then applied in the horizontal and vertical directions \citep{sobel1968,gonzalezwoods2018}:
\begin{align*}
    G_x&=\mathrm{Sobel}(\mathcal{P}_{\mathrm{fill}},x),\\
    G_y&=\mathrm{Sobel}(\mathcal{P}_{\mathrm{fill}},y).
\end{align*}
The gradient magnitude was
\begin{equation}
    \norm{\nabla\mathcal{P}}
    =\sqrt{G_x^2+G_y^2}.
    \label{eq:gradient}
\end{equation}
Finally, the original valid-data mask was restored, confining the gradient to the observed forest domain.

\section{Results}
\label{sec:results}

Applying the \HIAR{} estimator to the Sentinel-2 series from Mata de Santa Genebra produced a consolidated table of pixel-level parameters and a set of derived raster surfaces. Table~\ref{tab:empirical_summary} summarizes the principal empirical results.

\begin{table}[htbp]
\centering
\caption{Summary of the \HIAR{} application to the Mata de Santa Genebra ARIE.}
\label{tab:empirical_summary}
\begin{tabular}{lr}
\toprule
Metric & Value \\
\midrule
Valid pixels processed & 30,824 \\
Observation period & 2020--2023 \\
Observed processing time & approximately 10 min \\
Successful numerical terminations & 98.73\% \\
Median $\norm{\Phihat}$ & 0.6248 \\
Mean $\norm{\Phihat}$ & 0.6400 \\
Pixels with $\norm{\Phihat}\geq0.95$ & 3,872 (12.56\%) \\
Median B8 RMSE & 5.1551 percentage points \\
\bottomrule
\end{tabular}
\end{table}

The 98.73\% rate of successful numerical terminations reported by L-BFGS-B indicates predominantly stable operational behavior in the large-scale fitting exercise, although it does not guarantee a global optimum for every pixel. Summary statistics and raster surfaces were computed from all 30,824 finite estimates returned by the pipeline. For the remaining 1.27\%, the last admissible vector returned by the optimizer was retained after the radial projection implemented in the estimation routine, but without a formal convergence certificate. The mean and median of $\norm{\Phihat}$ were 0.6400 and 0.6248, respectively, and 3,872 pixels (12.56\%) met the operational high-persistence threshold.

\subsection{Temporal persistence and resilience indicator}

The \HIAR{} application produced a spatial surface of $\norm{\Phihat}$, which represents the overall temporal persistence of the multispectral process. Large values indicate that deviations tend to be retained for longer periods, whereas small values indicate faster temporal dissipation. In satellite vegetation applications, autoregressive persistence measures have been used as indirect indicators of slow recovery and possible resilience loss \citep{verbesselt2016,boulton2022,forzieri2022}.

\begin{figure}[htbp]
    \centering
    \includegraphics[width=0.95\textwidth]{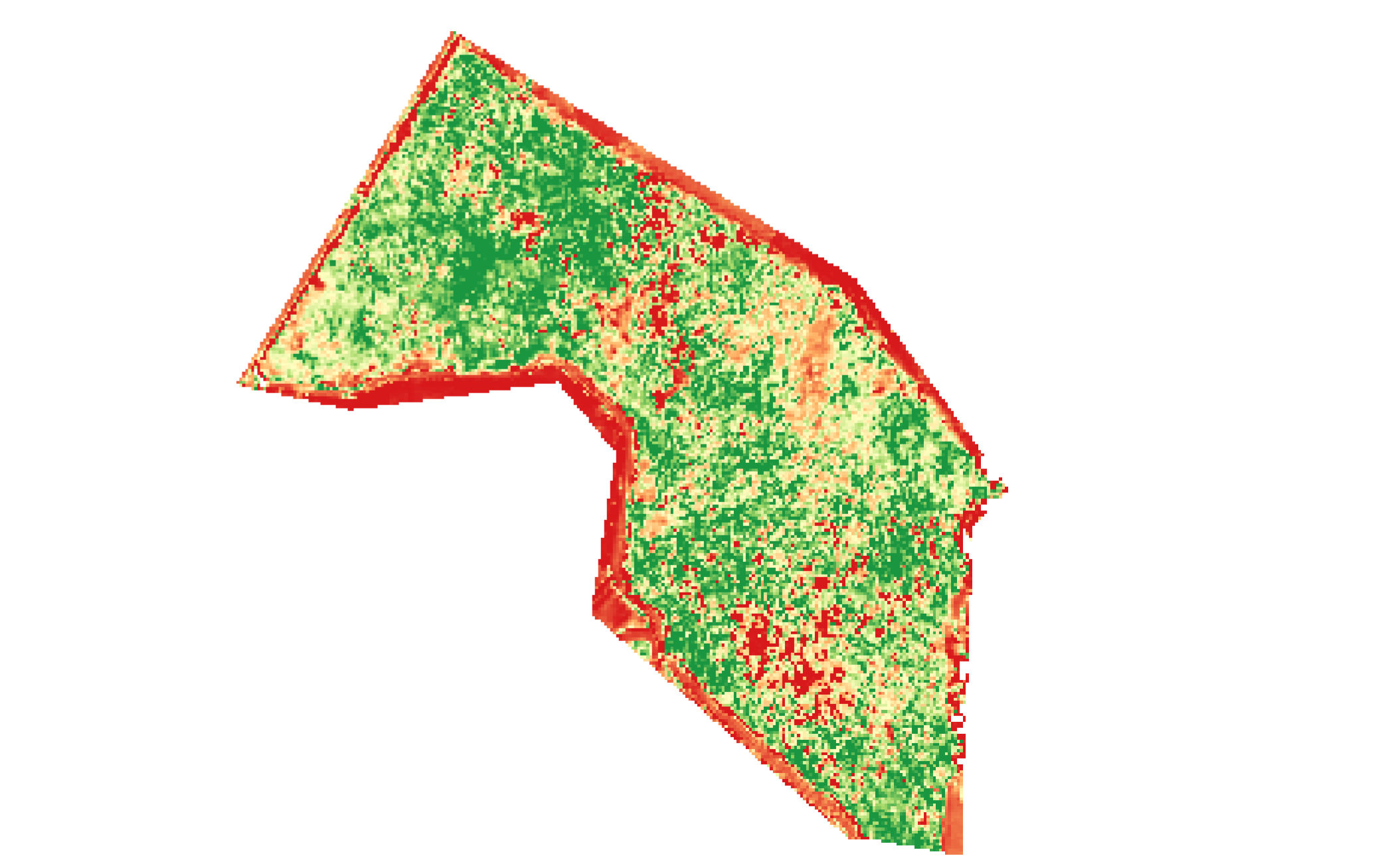}
    \caption{Pixel-level multispectral temporal persistence $\norm{\Phihat}$. Larger values indicate slower dissipation of multispectral deviations.}
    \label{fig:persistence}
\end{figure}
\FloatBarrier

The ecological interpretation of $\norm{\Phihat}$ should be restricted to an indirect temporal-persistence indicator rather than a direct physiological measurement. Where $\norm{\Phihat}$ is high, the pattern is compatible with slower post-disturbance recovery and potentially lower resilience; where it is low, the pattern is compatible with faster dissipation of spectral variation. This association alone cannot diagnose stress, degradation, or proximity to a critical transition.

For mapping purposes, an operational high-persistence mask was also defined by
\[
    \norm{\Phihat}\geq0.95.
\]
This threshold is an exploratory spatial-classification criterion, not a universal ecological boundary. In the Mata de Santa Genebra application, 3,872 pixels, corresponding to 12.56\% of valid pixels, met this operational criterion.

\subsection{Vector dominance}

\begin{figure}[htbp]
    \centering
    \includegraphics[width=0.95\textwidth]{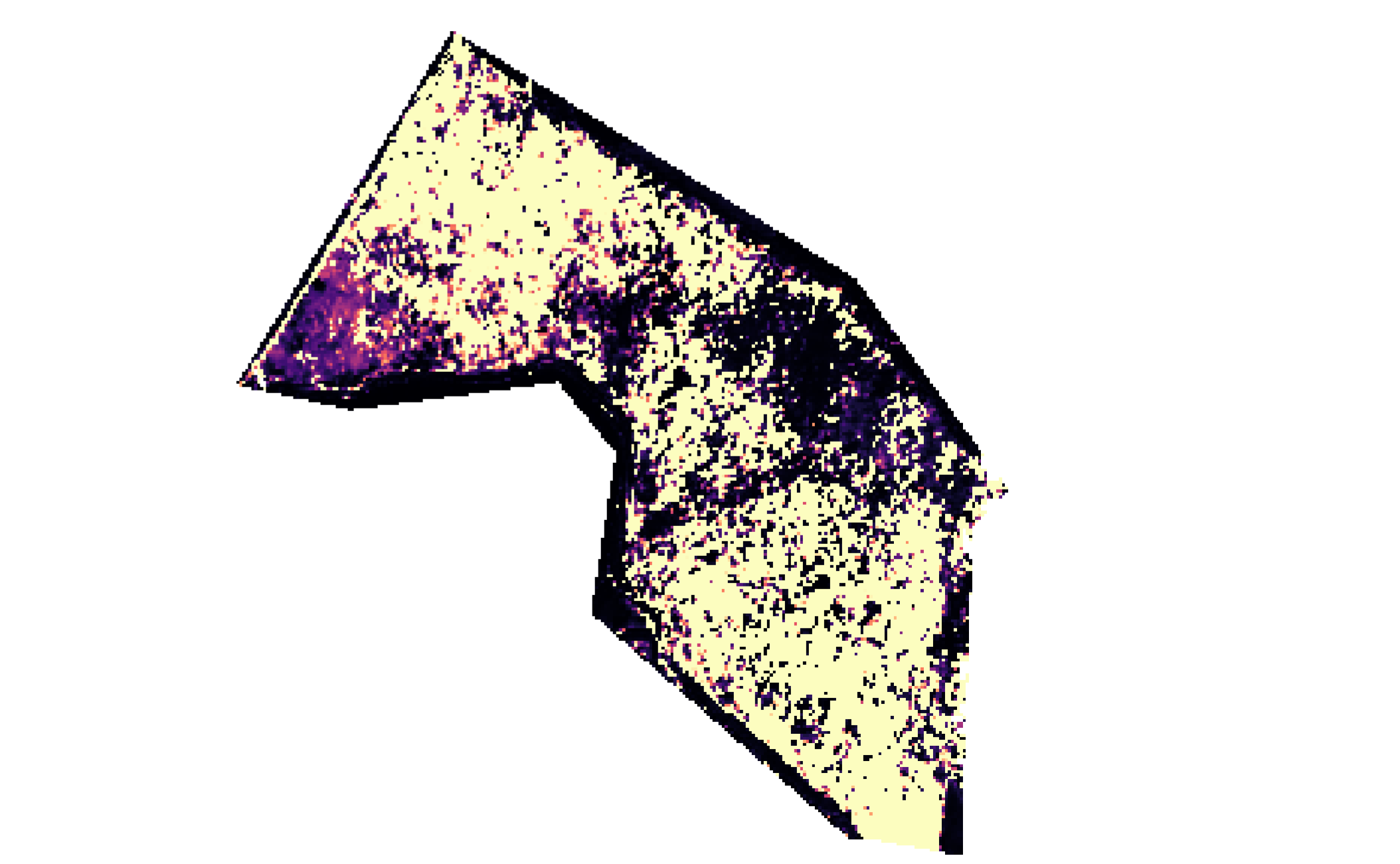}
    \caption{Vector dominance $\mathcal{D}$. Larger values indicate greater relative participation of the vector components of the quaternion parameter.}
    \label{fig:vector_dominance}
\end{figure}

Figure~\ref{fig:vector_dominance} shows the vector-dominance measure in~\eqref{eq:vector_dominance}. This measure summarizes the relative participation of the imaginary quaternion components, which enter the rotational part of the transition and produce structured mixing among the four observed components.

\FloatBarrier

Large $\mathcal{D}$ values indicate greater vector participation relative to the scalar component. This is compatible with more pronounced rotational dynamics among components, but it is not a general measure of correlation, causality, lagged dependence, or a physical interband mechanism.

\subsection{Predictive error}

One-step-ahead validation produced band-specific RMSE maps. Band B8 was emphasized because of its relationship with leaf and canopy structure and its comparatively large radiometric amplitude over vegetation \citep{knipling1970,jacquemoud2009,esa2015}.

\begin{figure}[htbp]
    \centering
    \includegraphics[width=0.95\textwidth]{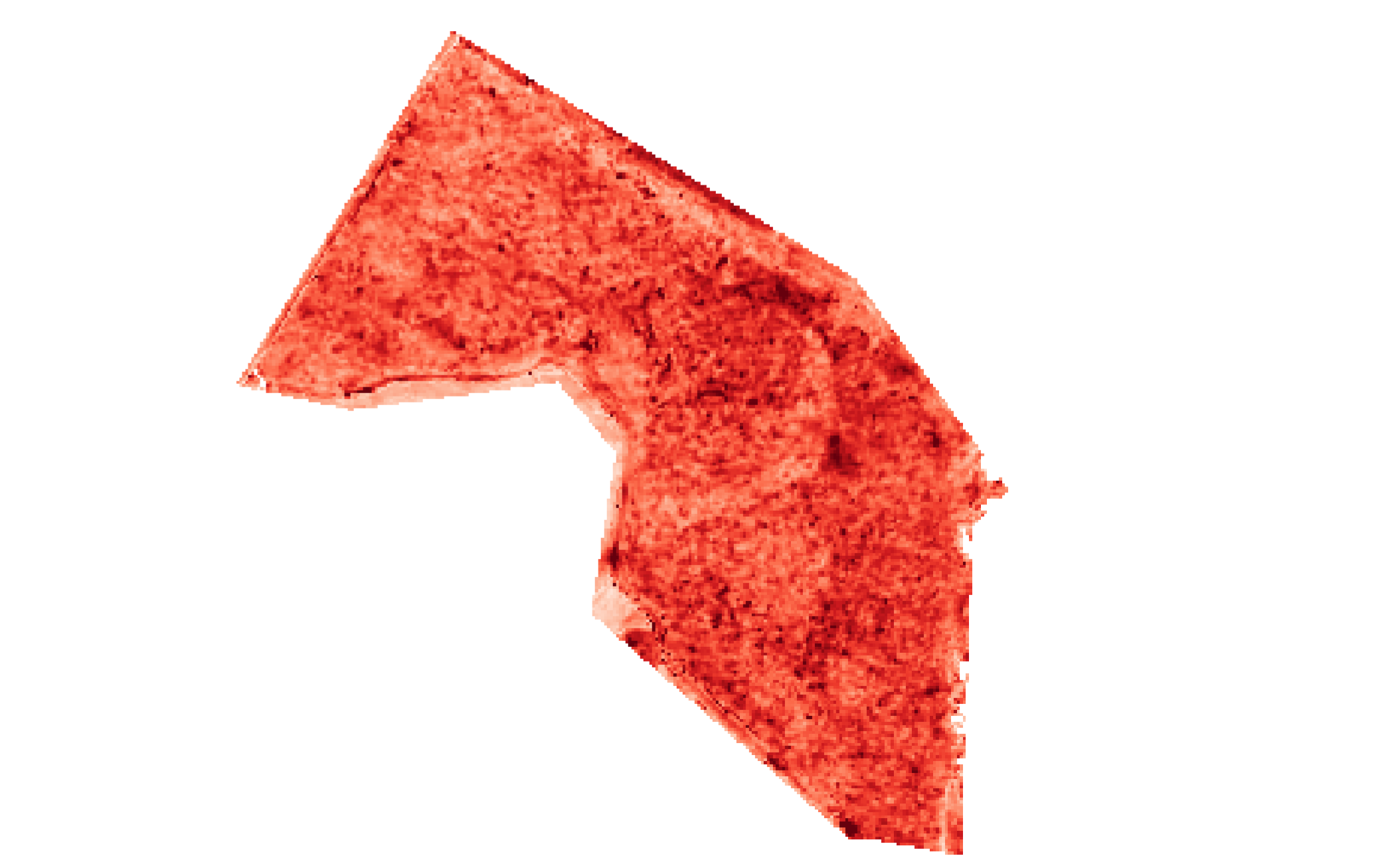}
    \caption{Predictive RMSE for band B8, computed on the chronologically held-out test segment.}
    \label{fig:rmse}
\end{figure}
\FloatBarrier

The median B8 RMSE was 5.1551 percentage points for the residual dynamics, using the percentage scale of the bands and a chronological training--test split. Because the same deterministic component can be added back to both prediction and observation, this numerical error is equivalent to that of the reconstructed series conditional on the fitted deterministic component. Low-RMSE regions indicate greater local ability to predict the residual B8 dynamics. High-RMSE regions suggest more irregular residual behavior, unmodeled perturbations, or greater local sensitivity to exogenous noise.

\subsection{Morphological gradient and exploratory edge identification}

\begin{figure}[htbp]
    \centering
    \includegraphics[width=0.95\textwidth]{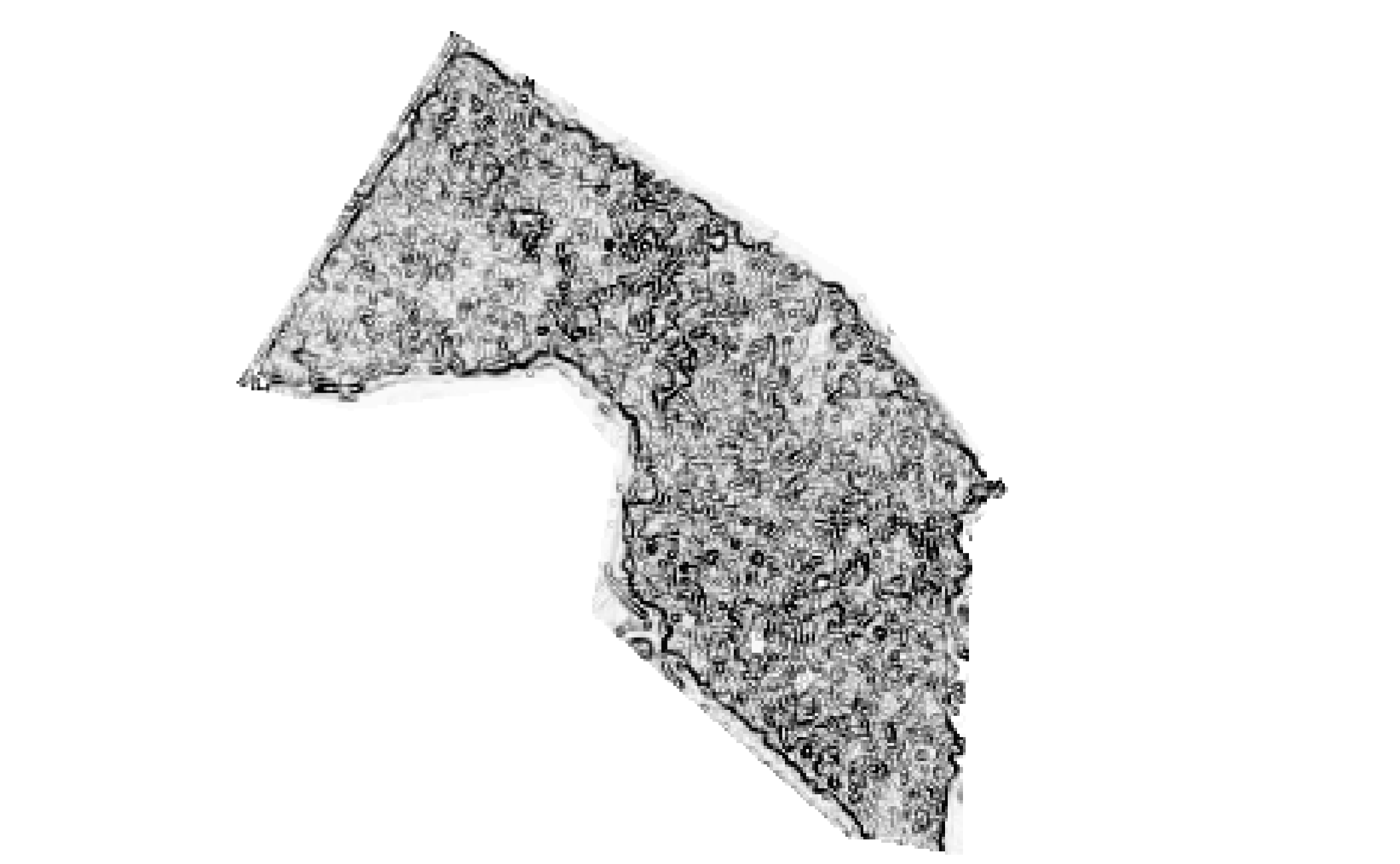}
    \caption{Morphological gradient of temporal persistence, computed with a Sobel operator after temporary nearest-neighbor filling and subsequent restoration of the original mask.}
    \label{fig:gradient}
\end{figure}

Figure~\ref{fig:gradient} shows the magnitude of the spatial gradient of temporal persistence. Unlike maps of absolute levels, the Sobel operator highlights abrupt transitions between neighboring pixels and yields an unsupervised descriptor of spatial discontinuities in estimated persistence \citep{sobel1968,gonzalezwoods2018}.

\FloatBarrier

Large $\norm{\nabla\mathcal{P}}$ values indicate abrupt local changes in temporal persistence. Such patterns may coincide with external edges, internal clearings, or structural transition zones, but may also reflect mixed pixels, differences in temporal data availability, or residual preprocessing effects. The gradient therefore identifies candidate edges rather than confirmed ecological classes. Interpretation is limited by the 10-m Sentinel-2 grid and the quality of the vegetation mask.

\section{Discussion}
\label{sec:discussion}

\subsection{Methodological contribution}

The \HIAR{} model extends the irregular autoregressive family \citep{eyheramendy2018,elorrieta2019,elorrieta2021} by representing four observed components through a single quaternion parameter. The resulting transition combines radial contraction and cross-component rotation \citep{ward1997,kuipers1999}. In multispectral applications, this structure avoids treating bands as independent processes or reducing them to isolated pairs.

The state-space representation permits estimation through the Gaussian innovation likelihood of the Kalman filter \citep{kalman1960,durbin2012,elorrieta2019}, conditional on the adopted plug-in covariance specifications. It accommodates irregular observations without temporal interpolation. The approximate diagonal specification for $Q_{t_j}$ was operationally stable in anisotropic multispectral data, preserved empirical marginal variances, and avoided positive-semidefiniteness failures during optimization.

\subsection{Ecological interpretation}

The norm $\norm{\Phihat}$ can be interpreted as a measure of multispectral temporal persistence. This persistence is compatible with slow-recovery indicators used for systems near critical regimes \citep{scheffer2009,dakos2012,verbesselt2016,boulton2022}, but it should not be interpreted as direct evidence of an ecological transition. Satellite time series record aggregate radiometric responses influenced by canopy structure, moisture, illumination, viewing geometry, and residual atmospheric noise \citep{knipling1970,jacquemoud2009,kandasamy2013}.

In Mata de Santa Genebra, the mean and median $\norm{\Phihat}$ were 0.6400 and 0.6248, while 12.56\% of valid pixels met the operational high-persistence threshold. High persistence was therefore spatially concentrated rather than uniform across the fragment. Ecological interpretation of these concentrations requires spatial covariates, management history, and field validation. Because each pixel received one time-constant $\Phihat$ estimated over the training segment, the application maps persistence over the study period but does not estimate its temporal evolution. Establishing an early-warning signal would require rolling windows, time-varying parameters, or longitudinal comparison of recovery indicators.

Vector dominance is an exploratory measure of the relative participation of the imaginary quaternion components. It identifies locations where the structured rotational component of the estimated dynamics is more pronounced. The persistence gradient, in turn, highlights abrupt spatial transitions between neighboring pixels \citep{sobel1968,gonzalezwoods2018}.

Candidate-edge identification is thus not a supervised image-classification step; it follows from spatial variation in a parameter estimated from temporal dynamics. This distinguishes the proposed approach from purely radiometric edge detectors \citep{gonzalezwoods2018}: the discontinuity emerges from estimated stochastic persistence rather than only from instantaneous reflectance contrast. Independent spatial validation remains necessary before these transitions can be assigned an ecological interpretation.

\subsection{Limitations}

The study has several limitations. First, ecological validation is indirect: the maps are derived from Sentinel-2 series rather than contemporaneous field measurements. The association between high persistence, slow recovery, and resilience loss is therefore an interpretive hypothesis rather than direct physiological confirmation. Autoregressive indicators derived from satellite observations may also have limited sensitivity in high-biomass vegetation \citep{smith2023}. Moreover, estimating a single $\Phi$ per pixel does not support inference about temporal trends in resilience or establish an early-warning signal.

Second, the threshold $\norm{\Phihat}\geq0.95$ is operational and exploratory. It highlights high-persistence areas but is not a universal threshold for collapse or degradation. Vector dominance is likewise a study-specific descriptor: it is unbounded and sensitive to scalar components near zero. Assigning B2 to the scalar component and B3, B4, and B8 to the imaginary components is part of the model parameterization, and sensitivity to alternative band orderings was not assessed. Similarly, the morphological gradient identifies discontinuities and candidate edges but was not quantitatively compared with reference forest boundaries.

Third, results depend on the SCL mask and the temporal availability of valid observations \citep{gee_s2,kandasamy2013}. Because retention depends on the \texttt{SCL == 4} classification, missingness may be related to observed surface conditions. Classification errors, persistent shadows, mixed boundary pixels, and heterogeneous pixels can also affect the estimates.

Fourth, the approximate diagonal specification of $Q_{t_j}$ was adopted for computational stability and large-scale processing. Monte Carlo generation used $P_0=I_4$, for which the expression is exact; during estimation, however, $P_0$ was again computed from empirical marginal variances. The application involved unequal marginal variances and therefore used an approximate anisotropic extension. Future formulations may consider alternative dynamic-noise structures, including cross-covariances and hierarchical models.

Fifth, the observation covariance $R=4I_4$ was fixed \textit{a priori}, without joint estimation or sensitivity analysis. This specification assumes homoscedastic and uncorrelated measurement errors across bands, times, and pixels. The empirical results are conditional on the adopted uncertainty level.

Sixth, the polar representation has a branch singularity on the negative real semiaxis. When the scalar component is negative and the vector norm is below $10^{-12}$, the unit axis of the quaternion power is not uniquely defined. The implementation handles this extremely narrow region through a zero-transition convention. The true parameters used in the Monte Carlo scenarios do not belong to this case because scenarios with negative scalar components have nonzero vector parts. The influence of intermediate optimizer evaluations in this region was not assessed separately.

Seventh, the 1.27\% of fits without a formal numerical-success flag were retained in the raster surfaces whenever they produced finite estimates. These values are the last admissible vectors returned by the optimizer and are not necessarily invalid fits; nevertheless, their potential local effect on derived surfaces, particularly the spatial gradient, was not quantified.

Finally, observed processing time depends on the hardware, concurrent load on the shared server, and parallelization choices. The reported value of approximately 10 minutes should therefore be interpreted as an operational record rather than a universal measure of computational performance.

\section{Conclusion}
\label{sec:conclusion}

This paper introduced the Hypercomplex Irregular Autoregressive (\HIAR) model, a quaternion formulation for multivariate time series observed at irregular times. Four components are represented by a single quaternion, temporal transitions are computed through real powers of the quaternion parameter, and a real $4\times4$ matrix embeds the dynamics in state-space form.

Estimation through the Gaussian innovation likelihood of the Kalman filter accommodates irregular observations without temporal interpolation, conditional on the adopted plug-in covariance specifications. Monte Carlo experiments provided numerical evidence of parameter recovery in synthetic scenarios. The empirical application demonstrated computational feasibility at spatial scale, processing 30,824 valid Sentinel-2 pixel series from Mata de Santa Genebra, with 98.73\% of optimizer runs reporting successful numerical termination.

The derived maps show that \HIAR{} can produce descriptors of multispectral temporal persistence, vector participation in quaternion dynamics, predictive error, and abrupt spatial transitions. The parameter norm provides a stochastic-memory indicator compatible with the slow-recovery paradigm associated with \textit{critical slowing down}, but it is neither a direct measure of ecological resilience nor a temporal early-warning signal on its own. The morphological gradient of persistence showed potential for exploratory identification of candidate forest edges and internal heterogeneity, which require independent spatial validation.

Future work should validate the maps against field data, incorporate spatial covariance among neighboring pixels, evaluate the model in other biomes, and compare \HIAR{} with alternative vector and multivariate models. The proposed formulation provides a mathematical and computational basis for treating temporal irregularity as an explicit component of multispectral dynamics rather than as a preprocessing obstacle.

\section*{Data Availability Statement}

The Python implementation used in this study is available at \url{https://github.com/BGCF11/ic-h-iar}. The repository contains the core \HIAR{} estimator, Monte Carlo scripts, Google Earth Engine extraction and preprocessing routines, pixelwise regional processing, and generation of the cartographic products. The source Sentinel-2 observations are publicly available through the Google Earth Engine collection \texttt{COPERNICUS/S2\_SR\_HARMONIZED}; extraction requires an authenticated Google Earth Engine account. The processed empirical data and Monte Carlo outputs supporting the reported tables and figures will be supplied as Supporting Information with the submission.

\section*{Conflict of Interest Statement}

The authors declare no conflict of interest.

\section*{Funding Information}

This work was supported by the S\~ao Paulo Research Foundation (FAPESP), Brazil, under grants 2023/02538-0 and 2025/21329-8.

\section*{Acknowledgments}

The authors thank the Institute of Mathematics, Statistics and Scientific Computing (IMECC) at the University of Campinas (UNICAMP) for providing computational resources.

\appendix
\setcounter{table}{0}
\renewcommand{\thetable}{A\arabic{table}}

\section{Full Monte Carlo validation tables}
\label{app:montecarlo}

\begin{table}[H]
\centering
\caption{Componentwise Monte Carlo results for Case 1 $(+/+)$.}
\label{tab:mc_case1}
\resizebox{\textwidth}{!}{%
\begin{tabular}{llrrrrrr}
\toprule
$N$ & Parameter & True value & Mean estimate & Abs. bias & SD & Iterations / mean time (s) & Time/iter. (ms) \\
\midrule
30  & $a$ & 0.7000 & 0.6788 & 0.0212 & 0.0454 & 10.663 / 0.0153 & 1.4321 \\
30  & $b$ & 0.3000 & 0.2864 & 0.0136 & 0.0623 & 10.663 / 0.0153 & 1.4321 \\
30  & $c$ & 0.3000 & 0.2866 & 0.0134 & 0.0609 & 10.663 / 0.0153 & 1.4321 \\
30  & $d$ & 0.3000 & 0.2881 & 0.0119 & 0.0603 & 10.663 / 0.0153 & 1.4321 \\
\midrule
100 & $a$ & 0.7000 & 0.6936 & 0.0064 & 0.0209 & 9.955 / 0.0291 & 2.9243 \\
100 & $b$ & 0.3000 & 0.2966 & 0.0034 & 0.0304 & 9.955 / 0.0291 & 2.9243 \\
100 & $c$ & 0.3000 & 0.2961 & 0.0039 & 0.0296 & 9.955 / 0.0291 & 2.9243 \\
100 & $d$ & 0.3000 & 0.2945 & 0.0055 & 0.0306 & 9.955 / 0.0291 & 2.9243 \\
\midrule
300 & $a$ & 0.7000 & 0.6976 & 0.0024 & 0.0111 & 9.384 / 0.0814 & 8.6776 \\
300 & $b$ & 0.3000 & 0.2985 & 0.0015 & 0.0167 & 9.384 / 0.0814 & 8.6776 \\
300 & $c$ & 0.3000 & 0.2987 & 0.0013 & 0.0158 & 9.384 / 0.0814 & 8.6776 \\
300 & $d$ & 0.3000 & 0.2996 & 0.0004 & 0.0176 & 9.384 / 0.0814 & 8.6776 \\
\bottomrule
\end{tabular}%
}
\end{table}

\begin{table}[H]
\centering
\caption{Componentwise Monte Carlo results for Case 2 $(-/-)$.}
\label{tab:mc_case2}
\resizebox{\textwidth}{!}{%
\begin{tabular}{llrrrrrr}
\toprule
$N$ & Parameter & True value & Mean estimate & Abs. bias & SD & Iterations / mean time (s) & Time/iter. (ms) \\
\midrule
30  & $a$ & -0.7000 & -0.6640 & 0.0360 & 0.0695 & 13.177 / 0.0137 & 1.0407 \\
30  & $b$ & -0.3000 & -0.2837 & 0.0163 & 0.0472 & 13.177 / 0.0137 & 1.0407 \\
30  & $c$ & -0.3000 & -0.2818 & 0.0182 & 0.0551 & 13.177 / 0.0137 & 1.0407 \\
30  & $d$ & -0.3000 & -0.2831 & 0.0169 & 0.0508 & 13.177 / 0.0137 & 1.0407 \\
\midrule
100 & $a$ & -0.7000 & -0.6898 & 0.0102 & 0.0208 & 12.146 / 0.0376 & 3.0955 \\
100 & $b$ & -0.3000 & -0.2950 & 0.0050 & 0.0213 & 12.146 / 0.0376 & 3.0955 \\
100 & $c$ & -0.3000 & -0.2960 & 0.0040 & 0.0212 & 12.146 / 0.0376 & 3.0955 \\
100 & $d$ & -0.3000 & -0.2960 & 0.0040 & 0.0218 & 12.146 / 0.0376 & 3.0955 \\
\midrule
300 & $a$ & -0.7000 & -0.6964 & 0.0036 & 0.0116 & 11.413 / 0.1048 & 9.1810 \\
300 & $b$ & -0.3000 & -0.2985 & 0.0015 & 0.0121 & 11.413 / 0.1048 & 9.1810 \\
300 & $c$ & -0.3000 & -0.2980 & 0.0020 & 0.0122 & 11.413 / 0.1048 & 9.1810 \\
300 & $d$ & -0.3000 & -0.2981 & 0.0019 & 0.0119 & 11.413 / 0.1048 & 9.1810 \\
\bottomrule
\end{tabular}%
}
\end{table}

\begin{table}[H]
\centering
\caption{Componentwise Monte Carlo results for Case 3 $(-/+)$.}
\label{tab:mc_case3}
\resizebox{\textwidth}{!}{%
\begin{tabular}{llrrrrrr}
\toprule
$N$ & Parameter & True value & Mean estimate & Abs. bias & SD & Iterations / mean time (s) & Time/iter. (ms) \\
\midrule
30  & $a$ & -0.9000 & -0.8421 & 0.0579 & 0.1409 & 16.736 / 0.0197 & 1.1785 \\
30  & $b$ & 0.1500 & 0.1393 & 0.0107 & 0.0369 & 16.736 / 0.0197 & 1.1785 \\
30  & $c$ & 0.1500 & 0.1410 & 0.0090 & 0.0381 & 16.736 / 0.0197 & 1.1785 \\
30  & $d$ & 0.1500 & 0.1395 & 0.0105 & 0.0419 & 16.736 / 0.0197 & 1.1785 \\
\midrule
100 & $a$ & -0.9000 & -0.8877 & 0.0123 & 0.0137 & 16.827 / 0.0617 & 3.6640 \\
100 & $b$ & 0.1500 & 0.1484 & 0.0016 & 0.0098 & 16.827 / 0.0617 & 3.6640 \\
100 & $c$ & 0.1500 & 0.1482 & 0.0018 & 0.0100 & 16.827 / 0.0617 & 3.6640 \\
100 & $d$ & 0.1500 & 0.1482 & 0.0018 & 0.0097 & 16.827 / 0.0617 & 3.6640 \\
\midrule
300 & $a$ & -0.9000 & -0.8959 & 0.0041 & 0.0071 & 17.580 / 0.1929 & 10.9751 \\
300 & $b$ & 0.1500 & 0.1493 & 0.0007 & 0.0052 & 17.580 / 0.1929 & 10.9751 \\
300 & $c$ & 0.1500 & 0.1494 & 0.0006 & 0.0055 & 17.580 / 0.1929 & 10.9751 \\
300 & $d$ & 0.1500 & 0.1493 & 0.0007 & 0.0056 & 17.580 / 0.1929 & 10.9751 \\
\bottomrule
\end{tabular}%
}
\end{table}

\begin{table}[H]
\centering
\caption{Componentwise Monte Carlo results for Case 4 $(+/-)$.}
\label{tab:mc_case4}
\resizebox{\textwidth}{!}{%
\begin{tabular}{llrrrrrr}
\toprule
$N$ & Parameter & True value & Mean estimate & Abs. bias & SD & Iterations / mean time (s) & Time/iter. (ms) \\
\midrule
30  & $a$ & 0.9000 & 0.8785 & 0.0215 & 0.0293 & 13.972 / 0.0144 & 1.0322 \\
30  & $b$ & -0.1500 & -0.1461 & 0.0039 & 0.0332 & 13.972 / 0.0144 & 1.0322 \\
30  & $c$ & -0.1500 & -0.1480 & 0.0020 & 0.0337 & 13.972 / 0.0144 & 1.0322 \\
30  & $d$ & -0.1500 & -0.1454 & 0.0046 & 0.0336 & 13.972 / 0.0144 & 1.0322 \\
\midrule
100 & $a$ & 0.9000 & 0.8951 & 0.0049 & 0.0121 & 14.188 / 0.0427 & 3.0108 \\
100 & $b$ & -0.1500 & -0.1480 & 0.0020 & 0.0157 & 14.188 / 0.0427 & 3.0108 \\
100 & $c$ & -0.1500 & -0.1490 & 0.0010 & 0.0161 & 14.188 / 0.0427 & 3.0108 \\
100 & $d$ & -0.1500 & -0.1483 & 0.0017 & 0.0163 & 14.188 / 0.0427 & 3.0108 \\
\midrule
300 & $a$ & 0.9000 & 0.8981 & 0.0019 & 0.0069 & 13.594 / 0.1239 & 9.1163 \\
300 & $b$ & -0.1500 & -0.1494 & 0.0006 & 0.0088 & 13.594 / 0.1239 & 9.1163 \\
300 & $c$ & -0.1500 & -0.1496 & 0.0004 & 0.0089 & 13.594 / 0.1239 & 9.1163 \\
300 & $d$ & -0.1500 & -0.1494 & 0.0006 & 0.0092 & 13.594 / 0.1239 & 9.1163 \\
\bottomrule
\end{tabular}%
}
\end{table}

\clearpage
\bibliographystyle{plainnat}
\bibliography{references}

@article{scheffer2009,
  author  = {Scheffer, Marten and Bascompte, Jordi and Brock, William A. and Brovkin, Victor and Carpenter, Stephen R. and Dakos, Vasilis and Held, Hermann and van Nes, Egbert H. and Rietkerk, Max and Sugihara, George},
  title   = {Early-warning signals for critical transitions},
  journal = {Nature},
  year    = {2009},
  volume  = {461},
  number  = {7260},
  pages   = {53--59},
  doi     = {10.1038/nature08227}
}

@article{dakos2012,
  author  = {Dakos, Vasilis and Carpenter, Stephen R. and Brock, William A. and Ellison, Aaron M. and Guttal, Vishwesha and Ives, Anthony R. and K{\'e}fi, Sonia and Livina, Valerie and Seekell, David A. and van Nes, Egbert H. and Scheffer, Marten},
  title   = {Methods for detecting early warnings of critical transitions in time series illustrated using simulated ecological data},
  journal = {PLOS ONE},
  year    = {2012},
  volume  = {7},
  number  = {7},
  pages   = {e41010},
  doi     = {10.1371/journal.pone.0041010}
}

@article{verbesselt2016,
  author  = {Verbesselt, Jan and Umlauf, Nikolaus and Hirota, Marina and Holmgren, Milena and van Nes, Egbert H. and Herold, Martin and Zeileis, Achim and Scheffer, Marten},
  title   = {Remotely sensed resilience of tropical forests},
  journal = {Nature Climate Change},
  year    = {2016},
  volume  = {6},
  number  = {11},
  pages   = {1028--1031},
  doi     = {10.1038/nclimate3108}
}

@article{boulton2022,
  author  = {Boulton, Chris A. and Lenton, Timothy M. and Boers, Niklas},
  title   = {Pronounced loss of {Amazon} rainforest resilience since the early 2000s},
  journal = {Nature Climate Change},
  year    = {2022},
  volume  = {12},
  number  = {3},
  pages   = {271--278},
  doi     = {10.1038/s41558-022-01287-8}
}

@article{forzieri2022,
  author  = {Forzieri, Giovanni and Dakos, Vasilis and McDowell, Nate G. and Ramdane, Alkama and Cescatti, Alessandro},
  title   = {Emerging signals of declining forest resilience under climate change},
  journal = {Nature},
  year    = {2022},
  volume  = {608},
  number  = {7923},
  pages   = {534--539},
  doi     = {10.1038/s41586-022-04959-9}
}

@article{smith2023,
  author  = {Smith, Taylor and Boers, Niklas},
  title   = {Reliability of vegetation resilience estimates depends on biomass density},
  journal = {Nature Ecology \& Evolution},
  year    = {2023},
  volume  = {7},
  number  = {11},
  pages   = {1799--1808},
  doi     = {10.1038/s41559-023-02194-7}
}

@article{eyheramendy2018,
  author  = {Eyheramendy, Susana and Elorrieta, Felipe and Palma, Wilfredo},
  title   = {An irregular discrete time series model to identify residuals with autocorrelation in astronomical light curves},
  journal = {Monthly Notices of the Royal Astronomical Society},
  year    = {2018},
  volume  = {481},
  number  = {4},
  pages   = {4311--4322},
  doi     = {10.1093/mnras/sty2487}
}

@article{elorrieta2019,
  author  = {Elorrieta, Felipe and Eyheramendy, Susana and Palma, Wilfredo},
  title   = {Discrete-time autoregressive model for unequally spaced time-series observations},
  journal = {Astronomy \& Astrophysics},
  year    = {2019},
  volume  = {627},
  pages   = {A120},
  doi     = {10.1051/0004-6361/201935560}
}

@article{elorrieta2021,
  author  = {Elorrieta, Felipe and Eyheramendy, Susana and Palma, Wilfredo and Ojeda, Cesar},
  title   = {A novel bivariate autoregressive model for predicting and forecasting irregularly observed time series},
  journal = {Monthly Notices of the Royal Astronomical Society},
  year    = {2021},
  volume  = {505},
  number  = {1},
  pages   = {1105--1116},
  doi     = {10.1093/mnras/stab1216}
}

@article{rehfeld2011,
  author  = {Rehfeld, Kira and Marwan, Norbert and Heitzig, Jobst and Kurths, J{\"u}rgen},
  title   = {Comparison of correlation analysis techniques for irregularly sampled time series},
  journal = {Nonlinear Processes in Geophysics},
  year    = {2011},
  volume  = {18},
  number  = {3},
  pages   = {389--404},
  doi     = {10.5194/npg-18-389-2011}
}

@article{kandasamy2013,
  author  = {Kandasamy, S. and Baret, F. and Verger, A. and Neveux, P. and Weiss, M.},
  title   = {A comparison of methods for smoothing and gap filling time series of remote sensing observations: application to {MODIS} {LAI} products},
  journal = {Biogeosciences},
  year    = {2013},
  volume  = {10},
  number  = {6},
  pages   = {4055--4071},
  doi     = {10.5194/bg-10-4055-2013}
}

@article{zhuwoodcock2014,
  author  = {Zhu, Zhe and Woodcock, Curtis E.},
  title   = {Automated cloud, cloud shadow, and snow detection in multitemporal {Landsat} data: an algorithm designed specifically for monitoring land cover change},
  journal = {Remote Sensing of Environment},
  year    = {2014},
  volume  = {152},
  pages   = {217--234},
  doi     = {10.1016/j.rse.2014.06.012}
}

@article{gorelick2017,
  author  = {Gorelick, Noel and Hancher, Matt and Dixon, Mike and Ilyushchenko, Simon and Thau, David and Moore, Rebecca},
  title   = {{Google Earth Engine}: Planetary-scale geospatial analysis for everyone},
  journal = {Remote Sensing of Environment},
  year    = {2017},
  volume  = {202},
  pages   = {18--27},
  doi     = {10.1016/j.rse.2017.06.031}
}

@manual{esa2015,
  author       = {{European Space Agency}},
  title        = {{Sentinel-2 User Handbook}},
  organization = {European Space Agency},
  year         = {2015},
  note         = {ESA Standard Document, Issue 1, Revision 2, 24 July 2015},
  url          = {https://sentinels.copernicus.eu/documents/247904/685211/Sentinel-2_User_Handbook}
}

@misc{gee_s2,
  author       = {{Google Earth Engine}},
  title        = {{COPERNICUS/S2\_SR\_HARMONIZED}: {Sentinel-2} {MSI} Level-2A Surface Reflectance Harmonized},
  howpublished = {Google Earth Engine Data Catalog},
  year         = {2026},
  note         = {Accessed: 2026-06-25},
  url          = {https://developers.google.com/earth-engine/datasets/catalog/COPERNICUS_S2_SR_HARMONIZED}
}

@article{knipling1970,
  author  = {Knipling, Edward B.},
  title   = {Physical and physiological basis for the reflectance of visible and near-infrared radiation from vegetation},
  journal = {Remote Sensing of Environment},
  year    = {1970},
  volume  = {1},
  number  = {3},
  pages   = {155--159},
  doi     = {10.1016/S0034-4257(70)80021-9}
}

@article{jacquemoud2009,
  author  = {Jacquemoud, St{\'e}phane and Verhoef, Wout and Baret, Fr{\'e}d{\'e}ric and Bacour, C{\'e}dric and Zarco-Tejada, Pablo J. and Asner, Gregory P. and Fran{\c c}ois, C{\'e}cile and Ustin, Susan L.},
  title   = {{PROSPECT} + {SAIL} models: A review of use for vegetation characterization},
  journal = {Remote Sensing of Environment},
  year    = {2009},
  volume  = {113},
  number  = {Suppl. 1},
  pages   = {S56--S66},
  doi     = {10.1016/j.rse.2008.01.026}
}

@article{guirao2011,
  author  = {Guir{\~a}o, {\^A}ngela Cruz and Teixeira Filho, Jos{\'e}},
  title   = {Preserva{\c c}{\~a}o de um fragmento florestal urbano: estudo de caso: a {ARIE} Mata de Santa Genebra, Campinas-{SP}},
  journal = {GEOUSP Espa{\c c}o e Tempo (Online)},
  year    = {2011},
  volume  = {15},
  number  = {1},
  pages   = {147--158},
  doi     = {10.11606/issn.2179-0892.geousp.2011.74193}
}

@misc{icmbio_santa_genebra,
  author       = {{Instituto Chico Mendes de Conserva{\c c}{\~a}o da Biodiversidade}},
  title        = {{ARIE} Mata de Santa Genebra},
  howpublished = {Portal Gov.br},
  year         = {2026},
  note         = {Accessed: 2026-06-25},
  url          = {https://www.gov.br/icmbio/pt-br/assuntos/biodiversidade/unidade-de-conservacao/unidades-de-biomas/mata-atlantica/lista-de-ucs/arie-mata-de-santa-genebra}
}

@book{ward1997,
  author    = {Ward, J. P.},
  title     = {Quaternions and Cayley Numbers: Algebra and Applications},
  series    = {Mathematics and Its Applications},
  volume    = {403},
  publisher = {Springer},
  address   = {Dordrecht},
  year      = {1997},
  doi       = {10.1007/978-94-011-5768-1},
  isbn      = {9780792345138}
}

@book{kuipers1999,
  author    = {Kuipers, Jack B.},
  title     = {Quaternions and Rotation Sequences: A Primer with Applications to Orbits, Aerospace, and Virtual Reality},
  publisher = {Princeton University Press},
  address   = {Princeton, NJ},
  year      = {1999},
  isbn      = {9780691102986}
}

@article{mebius2005,
  author        = {Mebius, Johan E.},
  title         = {A matrix-based proof of the quaternion representation theorem for four-dimensional rotations},
  journal       = {arXiv preprint math/0501249},
  year          = {2005},
  eprint        = {math/0501249},
  archivePrefix = {arXiv},
  url           = {https://arxiv.org/abs/math/0501249}
}

@article{kalman1960,
  author  = {Kalman, Rudolf E.},
  title   = {A new approach to linear filtering and prediction problems},
  journal = {Journal of Basic Engineering},
  year    = {1960},
  volume  = {82},
  number  = {1},
  pages   = {35--45},
  doi     = {10.1115/1.3662552}
}

@book{durbin2012,
  author    = {Durbin, James and Koopman, Siem Jan},
  title     = {Time Series Analysis by State Space Methods},
  edition   = {2},
  publisher = {Oxford University Press},
  address   = {Oxford},
  year      = {2012},
  isbn      = {9780199641178}
}

@article{byrd1995,
  author  = {Byrd, Richard H. and Lu, Peihuang and Nocedal, Jorge and Zhu, Ciyou},
  title   = {A limited memory algorithm for bound constrained optimization},
  journal = {SIAM Journal on Scientific Computing},
  year    = {1995},
  volume  = {16},
  number  = {5},
  pages   = {1190--1208},
  doi     = {10.1137/0916069}
}

@article{zhu1997,
  author  = {Zhu, Ciyou and Byrd, Richard H. and Lu, Peihuang and Nocedal, Jorge},
  title   = {Algorithm 778: {L-BFGS-B}: {Fortran} subroutines for large-scale bound-constrained optimization},
  journal = {ACM Transactions on Mathematical Software},
  year    = {1997},
  volume  = {23},
  number  = {4},
  pages   = {550--560},
  doi     = {10.1145/279232.279236}
}

@article{wilson2018,
  author  = {Wilson, Barry T. and Knight, Joseph F. and McRoberts, Ronald E.},
  title   = {Harmonic regression of {Landsat} time series for modeling attributes from national forest inventory data},
  journal = {ISPRS Journal of Photogrammetry and Remote Sensing},
  year    = {2018},
  volume  = {137},
  pages   = {29--46},
  doi     = {10.1016/j.isprsjprs.2018.01.006}
}

@article{verbesselt2010,
  author  = {Verbesselt, Jan and Hyndman, Rob and Newnham, Glenn and Culvenor, Darius},
  title   = {Detecting trend and seasonal changes in satellite image time series},
  journal = {Remote Sensing of Environment},
  year    = {2010},
  volume  = {114},
  number  = {1},
  pages   = {106--115},
  doi     = {10.1016/j.rse.2009.08.014}
}

@book{gonzalezwoods2018,
  author    = {Gonzalez, Rafael C. and Woods, Richard E.},
  title     = {Digital Image Processing},
  edition   = {4},
  publisher = {Pearson Education},
  address   = {Harlow},
  year      = {2018},
  isbn      = {9781292223070}
}

@misc{sobel1968,    
  author       = {Sobel, Irwin and Feldman, Gary},
  title        = {A 3x3 isotropic gradient operator for image processing},
  howpublished = {Presented at the Stanford Artificial Intelligence Project},
  year         = {1968},
  note         = {Historical source commonly cited for the Sobel operator}
}

\end{document}